\documentclass[a4paper]{article}

\usepackage{INTERSPEECH2019}
\usepackage[colorlinks]{hyperref}
\usepackage{url}
\usepackage{multirow}
\newcommand{\bhline}[1]{\noalign{\hrule height #1}}

\graphicspath{{figs/}}

\title{StarGAN-VC2:\\
  Rethinking Conditional Methods for StarGAN-Based Voice Conversion}
\name{Takuhiro Kaneko, Hirokazu Kameoka, Kou Tanaka, Nobukatsu Hojo}
\address{NTT Communication Science Laboratories, NTT Corporation, Japan}
\email{takuhiro.kaneko.tb@hco.ntt.co.jp}

\begin{document}

\maketitle

\begin{abstract}
  Non-parallel multi-domain voice conversion (VC) is a technique for learning mappings among multiple domains without relying on parallel data. This is important but challenging owing to the requirement of learning multiple mappings and the non-availability of explicit supervision. Recently, StarGAN-VC has garnered attention owing to its ability to solve this problem only using a single generator. However, there is still a gap between real and converted speech. To bridge this gap, we rethink conditional methods of StarGAN-VC, which are key components for achieving non-parallel multi-domain VC in a single model, and propose an improved variant called StarGAN-VC2. Particularly, we rethink conditional methods in two aspects: training objectives and network architectures. For the former, we propose a source-and-target conditional adversarial loss that allows all source domain data to be convertible to the target domain data. For the latter, we introduce a modulation-based conditional method that can transform the modulation of the acoustic feature in a domain-specific manner. We evaluated our methods on non-parallel multi-speaker VC. An objective evaluation demonstrates that our proposed methods improve speech quality in terms of both global and local structure measures. Furthermore, a subjective evaluation shows that StarGAN-VC2 outperforms StarGAN-VC in terms of naturalness and speaker similarity.\footnote{The converted speech samples are provided at \url{http://www.kecl.ntt.co.jp/people/kaneko.takuhiro/projects/stargan-vc2/index.html}.}
\end{abstract}

\section{Introduction}
\label{sec:introduction}

Voice conversion (VC) is a technique for converting the non/para-linguistic information between source and target speech while preserving the linguistic information. VC has been studied intensively owing to its high potential for various applications, such as speaking aids~\cite{AKainSC2007,KNakamuraSC2012} and style~\cite{ZInanogluSC2009,TTodaTASLP2012} and pronunciation~\cite{TKanekoIS2017b} conversion.

One well-established approach to VC involves statistical methods based on Gaussian mixture models (GMMs)~\cite{YStylianouTASP1998,TTodaTASLP2007,EHelanderTASLP2010}, neural networks (NNs) (including restricted Boltzmann machines (RBMs)~\cite{LChenTASLP2014,TNakashikaIEICE2014}, feed forward NNs (FNNs)~\cite{SDesaiTASLP2010,SMohammadiSLT2014,OKeisukeAPSIPAASC2017}, recurrent NNs (RNNs)~\cite{TNakashikaIS2014,LSunICASSP2015}, convolutional NNs (CNNs)~\cite{TKanekoIS2017b}, attention networks~\cite{KTanakaICASSP2019,HKameokaArXiv2018b}, and generative adversarial networks (GANs)~\cite{TKanekoIS2017b}), and exemplar-based methods using non-negative matrix factorization (NMF)~\cite{RTakashimaIEICE2013,ZWuTASLP2014}.

Many VC methods (including the above-mentioned) are categorized as parallel VC, which learns a mapping using the training data of parallel utterance pairs. However, obtaining such data is often time-consuming or impractical. Moreover, even if such data are obtained, most VC methods rely on a time alignment procedure, which occasionally fails and requires other painstaking processes, i.e., careful pre-screening or manual correction.

As a solution, non-parallel VC has begun to be studied. Non-parallel VC, which is comparable to parallel VC, is generally quite challenging to achieve  owing to its disadvantageous training conditions. To mitigate this difficulty, several studies have used additional data (e.g., parallel utterance pairs among reference speakers~\cite{AMouchtarisTASLP2006,CHLeeIS2006,TTodaIS2006,DSaitoIS2011}) or extra modules (e.g., automatic speech recognition (ASR) modules~\cite{FLXieIS2016,YSaitoICASSP2018}). These additional data and extra modules are useful for simplifying training but require other costs for preparation. To avoid such additional costs, recent studies have introduced probabilistic deep generative models, such as an RBM~\cite{TNakashikaTASLP2016}, variational autoencoders (VAEs)~\cite{CHsuIS2017,HKameokaTASLP2019}), and GANs \cite{CHsuIS2017,TKanekoArXiv2017}. Among them, CycleGAN-VC~\cite{TKanekoArXiv2017} (published \cite{TKanekoEUSIPCO2018} and further improved \cite{TKanekoICASSP2019}) shows promising results by configuring CycleGAN~\cite{JYZhuICCV2017,ZYiICCV2017,TKimICML2017} with a gated CNN~\cite{YDauphinICML2017} and identity-mapping loss~\cite{YTaigmanICLR2017}.
This makes it possible to learn a sequence-based mapping function without relying on parallel data. With this improvement, CycleGAN-VC performs comparably to parallel VC~\cite{TTodaTASLP2007}.

Along with non-parallel VC, another practically important issue is non-parallel multi-domain VC, i.e., learning mappings among multiple domains (e.g., multiple speakers) without relying on parallel data. This problem is challenging in terms of scalability because typical VC methods (including CycleGAN-VC) are designed to learn a one-to-one mapping; therefore, they require the learning of multiple generators to achieve multi-domain VC. For this problem, StarGAN-VC~\cite{HKameokaSLT2018} provides a promising solution by extending CycleGAN-VC to a conditional setting and incorporating domain codes. Through this extension, StarGAN-VC makes it possible to achieve non-parallel multi-domain VC by only using a single generator while maintaining the advantage of CycleGAN-VC. The subjective evaluation~\cite{HKameokaSLT2018} demonstrates that StarGAN-VC outperforms another state-of-the-art method, i.e., VAE/GAN-VC~\cite{CHsuIS2017}.

\begin{figure}[t]
  \centering
  \centerline{\includegraphics[width=\columnwidth]{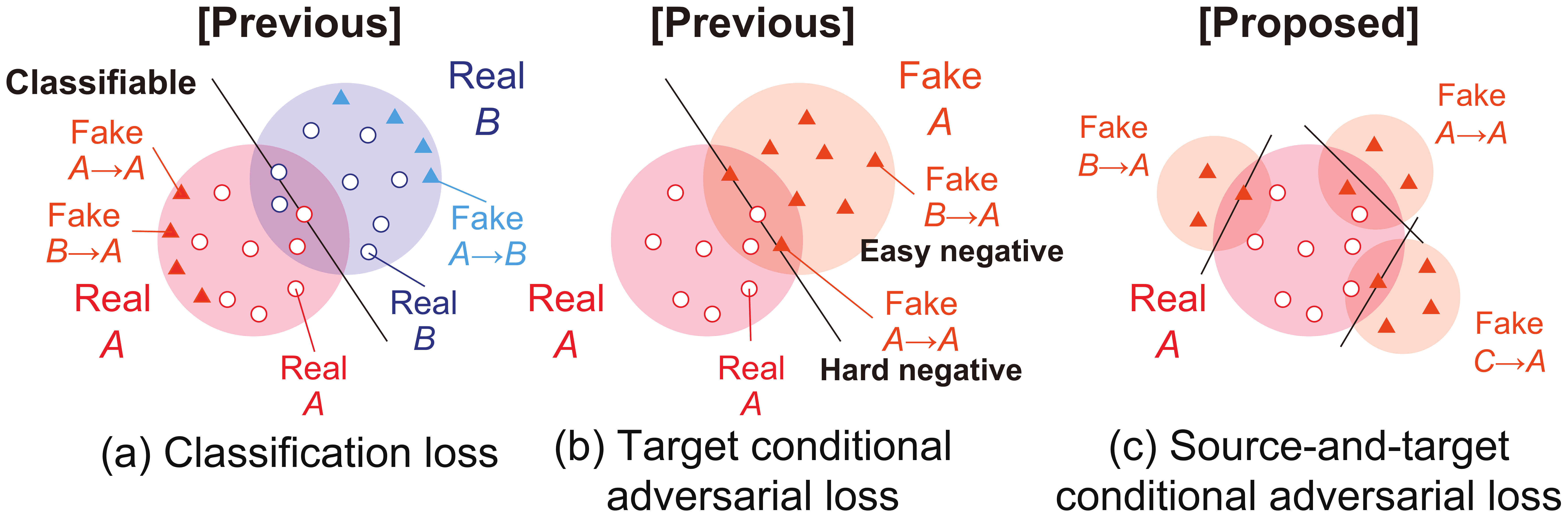}}
  \vspace{-2mm}
  \caption{Comparison of conditional methods in training objectives. ``$A$'' and ``$B$'' denote the domain codes and ``$A \rightarrow B$'' represents the data converted from ``$A$'' to ``$B$.'' Circle and triangle markers denote real and fake data, respectively. (a) In the classification loss, $G$ prefers to generate classifiable (i.e., far from the decision boundary) data. (b) In the target conditional adversarial loss, $D$ needs to simultaneously handle hard negative samples (e.g., conversion between the same speaker $A \rightarrow A$) and easy negative samples (e.g., conversion between completely different speakers $B \rightarrow A$). (c) The proposed source-and-target conditional adversarial loss can bring all the converted data close to the target data in both source-wise and target-wise manners.}    
  \label{fig:objective}
  \vspace{-4mm}
\end{figure}

However, even using StarGAN-VC, there is still an insurmountable gap between real and converted speech. To bridge this gap, we rethink conditional methods of StarGAN-VC, which are key components for solving non-parallel multi-domain VC using a single generator, and propose an improved variant called StarGAN-VC2. In particular, we rethink conditional methods in two aspects: training objectives and network architectures. For the former, we propose a source-and-target conditional adversarial loss, which encourages all source domain data to be converted into the target data. For the latter, we introduce a modulation-based conditional method that can transform the modulation of acoustic features in a domain-dependent manner. We examined the performance of the proposed methods on the multi-speaker VC task using the Voice Conversion Challenge 2018 (VCC 2018) dataset~\cite{VCC2018}. An objective evaluation demonstrates that the proposed methods effectively bring the converted acoustic feature sequence close to the target one in terms of both global and local structure measures. A subjective evaluation shows that StarGAN-VC2 outperforms StarGAN-VC in terms of both naturalness and speaker similarity.

In Section~\ref{sec:stargan-vc}, we review the conventional StarGAN-VC. In Section~\ref{sec:stargan-vc2}, we describe the proposed StarGAN-VC2. In Section~\ref{sec:experiments}, we report the experimental results. We conclude in Section~\ref{sec:conclusions} with a brief summary and mention of future work.

\section{Conventional StarGAN-VC}
\label{sec:stargan-vc}

\subsection{Training objectives}
\label{subsec:stargan-vc-obj}

The aim of StarGAN-VC is to obtain a single generator $G$ that learns mappings among multiple domains (e.g., multiple speakers). To achieve this, StarGAN-VC extends CycleGAN-VC to a conditional setting with a domain code (e.g., a speaker identifier). More precisely, StarGAN-VC learns a generator $G$ that converts an input acoustic feature ${\bm x}$ into an output feature ${\bm x}'$ conditioned on the target domain code $c'$, i.e., $G({\bm x}, c') \rightarrow {\bm x}'$. Here, let ${\bm x} \in \mathbb{R}^{Q \times T}$ be an acoustic feature sequence where $Q$ is the feature dimension and $T$ is the sequence length, and let $c \in \{ 1, \dots, N \}$ be the corresponding domain code where $N$ is the number of domains. Inspired by StarGAN~\cite{YChoiCVPR2018}, which was originally proposed in computer vision for multi-domain image-to-image translation, StarGAN-VC solves this problem by using an adversarial loss~\cite{IGoodfellowNIPS2014}, classification loss~\cite{AOdenaICML2017}, and cycle-consistency loss~\cite{TZhouCVPR2016}. Additionally, inspired by CycleGAN-VC~\cite{TKanekoArXiv2017}, StarGAN-VC also uses an identity-mapping loss~\cite{YTaigmanICLR2017} to preserve linguistic composition.

\textbf{Adversarial loss:}
The adversarial loss is used to render the converted feature indistinguishable from the real target feature:
\begin{flalign}
  \label{eqn:adv}
  & {\cal L}_{t\text{-}adv} = \mathbb{E}_{({\bm x}, c) \sim P({\bm x}, c)}
  [\log D({\bm x}, c)]
  \nonumber \\
  & \:\:\:\:\:\:\:\:\: + \mathbb{E}_{{\bm x} \sim P({\bm x}), c' \sim P(c')} [\log (1 - D(G({\bm x}, c'), c'))],
\end{flalign}
where $D$ is a target conditional discriminator~\cite{MMirzaArXiv2014}. By maximizing this loss, $D$ attempts to learn the best decision boundary between the converted and real acoustic features conditioned on the target domain codes ($c$ and $c'$). In contrast, $G$ attempts to render the converted feature indistinguishable from real acoustic features conditioned on $c'$ by minimizing this loss.

\textbf{Classification loss:}
The aim of StarGAN-VC is to synthesize the acoustic feature that belongs to the target domain. To achieve this, the classification loss is used. First, the classifier $C$ is trained for real acoustic features:
\begin{flalign}
  \label{eqn:cls_r}
  {\cal L}_{cls}^r = \mathbb{E}_{({\bm x}, c) \sim P({\bm x}, c)} [- \log C(c | {\bm x})],
\end{flalign}
where $C$ attempts to classify a real acoustic feature ${\bm x}$ to the corresponding domain $c$ by minimizing this loss. Subsequently, $G$ is optimized for $C$:
\begin{flalign}
  \label{eqn:cls_f}
  {\cal L}_{cls}^f = \mathbb{E}_{{\bm x} \sim P({\bm x}), c' \sim P(c')} [- \log C(c' | G({\bm x}, c'))],
\end{flalign}
where $G$ attempts to generate an acoustic feature that is classified to the target domain $c'$ by minimizing this loss.

\textbf{Cycle-consistency loss:}
Although the adversarial loss and classification loss encourage a converted acoustic feature to become realistic and classifiable, respectively, they do not guarantee that the converted feature will preserve the input composition. To mitigate this problem, the cycle-consistency loss is used:
\begin{flalign}
  \label{eqn:cyc}
  {\cal L}_{cyc} = \mathbb{E}_{({\bm x}, c) \sim P({\bm x}, c), c' \sim P(c')} [ \| {\bm x} - G(G({\bm x}, c'), c) \|_1 ].
\end{flalign}
This cyclic constraint encourages $G$ to find out an optimal source and target pair that does not compromise the composition.

\textbf{Identity-mapping loss:}
To impose a further constraint on the input preservation, the identity-mapping loss is used:
\begin{flalign}
  \label{eqn:id}
  {\cal L}_{id} = \mathbb{E}_{({\bm x}, c) \sim P({\bm x}, c)} [ \| G({\bm x}, c) - {\bm x} \| ].
\end{flalign}

\textbf{Full objective:}
The full objective is written as
\begin{flalign}
  {\cal L}_D = & \: - {\cal L}_{t\text{-}adv},
  \\
  {\cal L}_C = & \: \lambda_{cls} {\cal L}_{cls}^r,
  \\
  {\cal L}_G = & \: {\cal L}_{t\text{-}adv} + \lambda_{cls} {\cal L}_{cls}^f + \lambda_{cyc} {\cal L}_{cyc} + \lambda_{id} {\cal L}_{id},
\end{flalign}
where $D$, $C$, and $G$ are optimized by minimizing ${\cal L}_D$, ${\cal L}_C$, and ${\cal L}_G$, respectively.

\begin{figure}[t]
  \centering
  \centerline{\includegraphics[width=1\columnwidth]{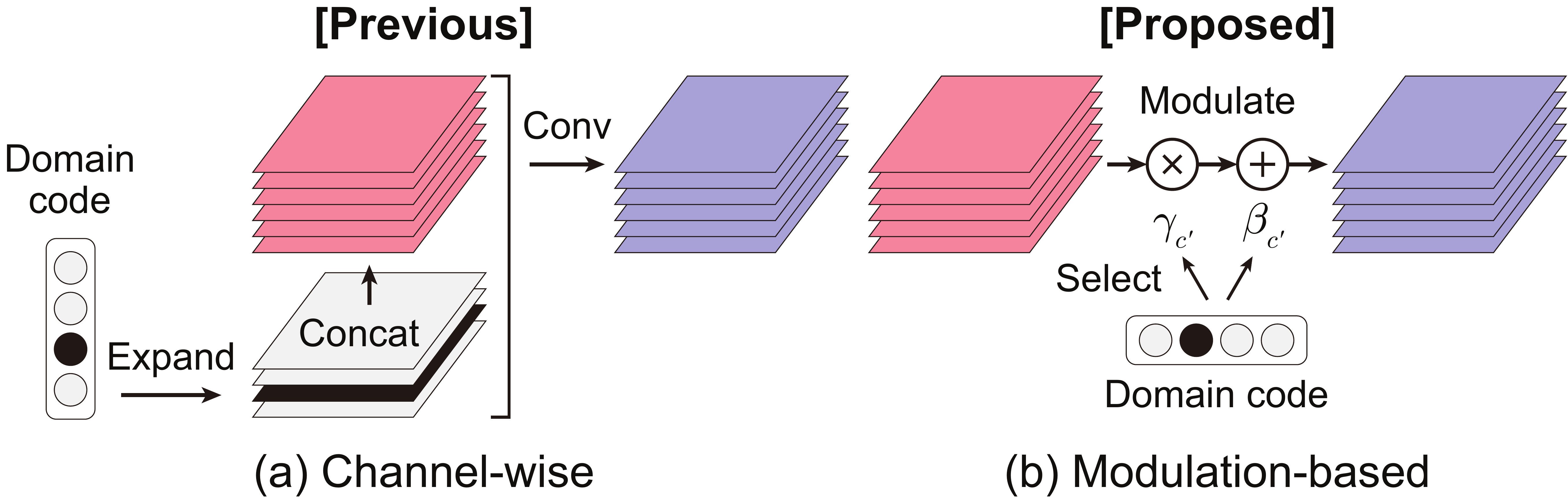}}
  \vspace{-2mm}
  \caption{Comparison of conditional methods in generator networks. We consider the case when convolutional networks are used. Such networks are commonly used in state-of-the-art VC models (e.g., CycleGAN-VC~\cite{TKanekoArXiv2017}and StarGAN-VC~\cite{HKameokaSLT2018}).}
  \label{fig:network}
  \vspace{-4mm}
\end{figure}

\subsection{Network architectures}
\label{subsec:stargan-vc-arch}

Regarding the network architectures, this study focuses on the conditional method in the generator. Hence, here we review the StarGAN-VC generator network architecture. As shown in Figure~\ref{fig:network}(a), StarGAN-VC incorporates conditional information into the generator in a channel-wise manner, i.e., first creates the one-hot vector indicating the domain code, subsequently expands the one-hot vector to the feature map size, and finally concatenates it to the feature map. Concatenated features are convoluted together and propagated to the next layer.

\section{StarGAN-VC2}
\label{sec:stargan-vc2}

\subsection{Rethinking conditional method in training objectives}
\label{subsec:stargan-vc2-obj}

We first rethink a conditional method in training objectives. As described in Section~\ref{subsec:stargan-vc-obj}, StarGAN-VC uses two conditional methods to make the converted feature belonging to the target domain: the classification loss (Equations~\ref{eqn:cls_r} and \ref{eqn:cls_f}) and the target conditional adversarial loss (Equation~\ref{eqn:adv}). We illustrate their training strategies in Figure~\ref{fig:objective}(a) and (b), respectively.

In the classification loss (Figure~\ref{fig:objective}(a)), via Equation~\ref{eqn:cls_r}, the decision boundary (black line) is learned among real-data domains (e.g., between ``Real $A$'' and ``Real $B$'' in Figure~\ref{fig:objective}(a)). For this decision boundary, $G$ attempts to generate easily ``classifiable'' data via Equation~\ref{eqn:cls_f}. This means that $G$ prefers to generate data that are far from the decision boundary even when the real data exist around the decision boundary. As discussed elsewhere~\cite{AOdenaICML2017,TMiyatoICLR2018,TKanekoBMVC2019}, this prevents $G$ from covering the whole real data distribution. In VC, this may result in a partial conversion.

Meanwhile, the target conditional adversarial loss (Figure~\ref{fig:objective}(b)) encourages the generated data close to the real data conditioned on the target domain code. As discussed in the previous study~\cite{TMiyatoICLR2018}, this objective prevents $G$ from leaning towards generating only classifiable data. However, a possible difficulty is that this loss needs to simultaneously handle diverse data, including hard negative samples (e.g., conversion between the same speaker $A \rightarrow A$ in Figure~\ref{fig:objective}(b)) and easy negative samples (e.g, conversion between completely different speakers $B \rightarrow A$ in Figure~\ref{fig:objective}(b)). This unfair condition makes it difficult to bring all the converted data close to real target data.

To solve this problem, we develop a source-and-target conditional adversarial loss defined as
\begin{flalign}
  \label{eqn:stadv}
  & {\cal L}_{st\text{-}adv} = \mathbb{E}_{({\bm x, c}) \sim P({\bm x}, c), c' \sim P(c')} [\log D({\bm x}, c', c)]
  \nonumber \\
  & \:\:\:\:\:\:\:\:\: + \mathbb{E}_{({\bm x, c}) \sim P({\bm x}, c), c' \sim P(c')} [\log D(G({\bm x}, c, c'), c, c')],
\end{flalign}
where $c' \sim P(c')$ is randomly sampled independently of real data. Differently from Equation~\ref{eqn:adv}, both $G$ and $D$ are conditioned on the source code $c'$ in addition to the target code $c$. We call such $G$ and $D$ a source-and-target conditional generator and discriminator, respectively. As shown in Figure~\ref{fig:objective}(c), by using both source and target domain codes as conditional information, this loss encourages all the converted data to be close to real data in both source-wise and target-wise manners. This resolves the unfair training condition in the target conditional adversarial loss (Figure~\ref{fig:objective}(b)) and allows all the source domain data to be converted into the target domain data.

One possible disadvantage of the source-and-target conditional generator is that this requires the availability of the source code in inference, which is not required in the conventional StarGAN-VC. However, note that speaker recognition has been actively studied (e.g., \cite{MMclarenICASSP2015}), and this problem can be alleviated by using it as a pre-process.

\subsection{Rethinking conditional method in networks}
\label{subsec:stargan-vc2-arch}

As indicated by previous studies on VC postfilters (e.g., global variance~\cite{TTodaTASLP2007} and modulation spectrum~\cite{STakamichiICASSP2014} postfilters), accurate modulation translation is important to achieve high-quality VC. Particularly, to achieve multi-domain VC only using a single generator, a framework must be incorporated that can conduct diverse domain-specific modulations effectively. For this challenge, a channel-wise conditional method (Figure~\ref{fig:network}(a)) is not effective because the concatenated conditional information can be additively used in a convolutional procedure but cannot be multicatively used to modulate features. To alleviate this problem, we introduce a modulation-based conditional method, which can directly modulate features in a domain-dependent manner. In particular, we introduce a conditional instance normalization (CIN)~\cite{VDumoulinICLR2017}, which was originally proposed in computer vision for style transfer. As shown in Figure~\ref{fig:network}(b), given the feature ${\bm f}$, CIN conducts the following procedure:
\begin{flalign}
  \label{eqn:cin}
  {\rm CIN}({\bm f}; c') = \gamma_{c'} \left( \frac{{\bm f} - \mu({\bm f})}{\sigma({\bm f})} \right) + \beta_{c'},
\end{flalign}
where $\mu({\bm f})$ and $\sigma({\bm f})$ are the average and standard deviation of ${\bm f}$ that are calculated over for each instance. $\gamma_{c'}$ and $\beta_{c'}$ are domain-specific scale and bias parameters that allow the modulation to be transformed in a domain-specific manner. These parameters are learnable and optimized through training.

In the above, we explain the case when the generator is conditioned on the target domain code $c'$. When using a source-and-target conditional generator introduced in Equation~\ref{eqn:stadv}, we replace $\gamma_{c'}$ and $\beta_{c'}$ with $\gamma_{c, c'}$ and $\beta_{c, c'}$, respectively, which are selected depending on both the source $c$ and target $c'$.

\section{Experiments}
\label{sec:experiments}

\begin{figure*}[t]
  \centering
  \centerline{\includegraphics[width=1\textwidth]{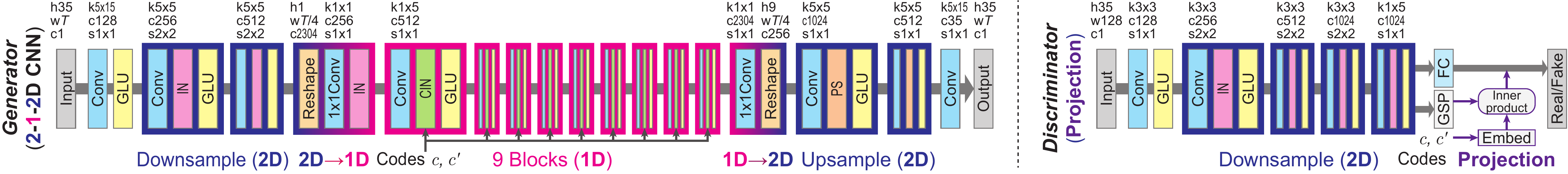}}
  \vspace{-2mm}
  \caption{Generator and discriminator network architectures. In input, output, and reshape layers, {\tt h}, {\tt w}, and {\tt c} represent height, width, and number of channels, respectively. In each convolution layer, {\tt k}, {\tt c}, and {\tt s} denote kernel size, number of channels, and stride, respectively. IN, GLU, PS, and GSP indicate instance normalization~\cite{DUlyanovArXiv2016}, gated linear unit~\cite{YDauphinICML2017}, pixel shuffler~\cite{WShiCVPR2016}, and global sum pooling, respectively. The generator is fully convolutional~\cite{JLongCVPR2015}. This allows an arbitrary length $T$ to be input in inference.}
  \label{fig:implementation}
  \vspace{-4mm}
\end{figure*}

\subsection{Experimental conditions}
\label{subsec:exp_conditions}

\textbf{Dataset:}
We evaluated our method on the multi-speaker VC task using VCC 2018~\cite{VCC2018}, which contains recordings of professional US English speakers. Following the StarGAN-VC study \cite{HKameokaSLT2018}, we selected a subset of speakers as covering all inter- and intra-gender conversions: VCC2SF1, VCC2SF2, VCC2SM1, and VCC2SM2, where F and M indicate female and male speakers, respectively. Thus, the number of domains $N$ is set to 4. Our goal is to learn $4 \times 3 = 12$ different source-and-target mappings in a single model. Each speaker has sets of 81 and 35 sentences for training and evaluation, respectively. The recordings were downsampled to 22.05 kHz for this challenge. We extracted 34 Mel-cepstral coefficients (MCEPs), logarithmic fundamental frequency ($\log F_0$), and aperiodicities (APs) every 5 ms by using the WORLD analyzer \cite{MMoriseIEICE2016}.

\textbf{Conversion process:}
In these experiments, we focused on analyzing the performance in MCEP conversion. Hence, we applied the proposed method only to MCEP conversion,\footnote{For reference, the converted speech samples, in which the proposed method was applied to convert all acoustic features (namely, MCEPs, band APs, continuous $\log F_0$, and voice/unvoice indicator), are provided at \url{http://www.kecl.ntt.co.jp/people/kaneko.takuhiro/projects/stargan-vc2/index.html}. Even in this challenging setting, StarGAN-VC2 works reasonably well.} and for the other parts, we used typical methods, i.e., converted $\log F_0$ using logarithm Gaussian normalized transformation~\cite{KLiuFSKD2007}, directly used APs, and synthesized speech using the WORLD vocoder~\cite{MMoriseIEICE2016}. To examine the pure effect of the proposed methods, we did not use any postfilter~\cite{TKanekoICASSP2017,TKanekoIS2017a,KTanakaSLT2018} or powerful vocoder such as the WaveNet vocoder~\cite{OAaronArXiv2016,ATamamoriIS2017}. Incorporating them remains possible future work.

\textbf{Implementation details:}
We designed the network architectures on the basis of CycleGAN-VC2~\cite{TKanekoICASSP2019}, i.e., we used a 2-1-2D CNN in $G$ and a 2D CNN in $D$. We formulate $D$ using the projection discriminator \cite{TMiyatoICLR2018}. In the pre-experiment, we found that skip connections in residual blocks~\cite{KHeCVPR2016} result in partial conversion. Thus, we removed them in $G$. The details of the network architectures are given in Figure~\ref{fig:implementation}. For a GAN objective, we used a least squares GAN \cite{XMaoICCV2017}. We conducted speaker-wise normalization for a pre-process. We trained the networks using the Adam optimizer \cite{DPKingmaICLR2015} with a batch size of 8, in which we used a randomly cropped segment (128 frames) as one instance. The number of iterations was set to $3 \times 10^5$, learning rates for $G$ and $D$ were set to 0.0002 and 0.0001, respectively, and the momentum term was set to 0.5. We set $\lambda_{cyc} = 10$, $\lambda_{id} = 5$, and  $\lambda_{cls} = 1$. We used ${\cal L}_{id}$ only for the first $10^4$ iterations to stabilize the training at the beginning.

\begin{table}[t]  
  \centering
  \caption{Comparison of MCD and MSD among models using different conditional methods in training objectives. We fix the conditional method in $G$ network as modulation-based.}
  \vspace{-2mm}
  \label{tab:comp_obj}
  \scalebox{0.9}{
    \begin{tabular}{l|cc}
      \bhline{1pt}
      Objective & MCD [dB] & MSD [dB]
      \\ \bhline{0.75pt}
      ${\cal L}_{cls}$
                    & 7.73 $\pm$ .07
                               & 1.96 $\pm$ .03
      \\
      ${\cal L}_{t\text{-}adv}$
                    & 7.21 $\pm$ .16
                               & 2.87 $\pm$ .51
      \\
      ${\cal L}_{t\text{-}adv} + {\cal L}_{cls}$ (StarGAN-VC)
                    & 7.11 $\pm$ .10
                               & 2.41 $\pm$ .13
      \\ \hline
      ${\cal L}_{st\text{-}adv}$ (StarGAN-VC2)
                    & \textbf{6.90 $\pm$ .07}
                               & \textbf{1.89 $\pm$ .03} 
      \\
      \bhline{1pt}
    \end{tabular}
  }
  \vspace{-2mm}
\end{table}

\begin{table}[t]  
  \centering  
  \caption{Comparison of MCD and MSD among models using different conditional methods in $G$ networks. We fix the conditional method in the training objective as
    ${\cal L}_{st\text{-}adv}$.}
  \vspace{-2mm}
  \label{tab:comp_net}
  \scalebox{0.9}{
    \begin{tabular}{l|cc}
      \bhline{1pt}
      $G$ network & MCD [dB] & MSD [dB]
      \\ \bhline{0.75pt}
      Channel-wise (StarGAN-VC)
                  & 6.90 $\pm$ .08
                             & 2.55 $\pm$ .20
      \\ \hline
      Modulation-based (StarGAN-VC2)
                  & 6.90 $\pm$ .07
                             & \textbf{1.89 $\pm$ .03} 
      \\
      \bhline{1pt}
    \end{tabular}
  }
  \vspace{-5mm}
\end{table}

\subsection{Objective evaluation}
\label{subsec:objective}

We conducted an objective evaluation to validate the advantages of the proposed conditional methods over other conditional methods. The same as the previous study \cite{TKanekoICASSP2019}, we used two evaluation metrics for comprehensive analysis: the Mel-cepstral distortion ({MCD}), which measures the global structural differences by calculating the distance between the target and converted MCEPs, and the modulation spectra distance ({MSD}), which measures the local structural differences by computing the distance between the target and converted logarithmic modulation spectra of MCEPs. For both metrics, a smaller value indicates that the target and converted features are more similar.

We conducted comparative studies in two aspects: training objectives and network architectures, which are listed in Tables \ref{tab:comp_obj} and \ref{tab:comp_net}, respectively. We have calculated the scores averaged over three models trained with different initializations to eliminate the effect of initialization. In Table \ref{tab:comp_obj}, the proposed source-and-target conditional loss ${\cal L}_{st\text{-}adv}$ outperforms the other losses in terms of both the MCD and MSD. This indicates that the proposed loss is useful for improving the feature quality in terms of both the global and local structure measures. In Table \ref{tab:comp_net}, the proposed modulation-based conditional method outperforms the conventional channel-wise conditional method in terms of the MSD. This indicates that the proposed architecture is particularly useful for improving the local structure. Through these experiments, we empirically confirm that the proposed conditional methods in objectives and networks effectively bring the converted acoustic feature sequence close to the target one.

\begin{figure}[t]
  \vspace{-1.9mm}
  \centering
  \centerline{\includegraphics[width=1\columnwidth]{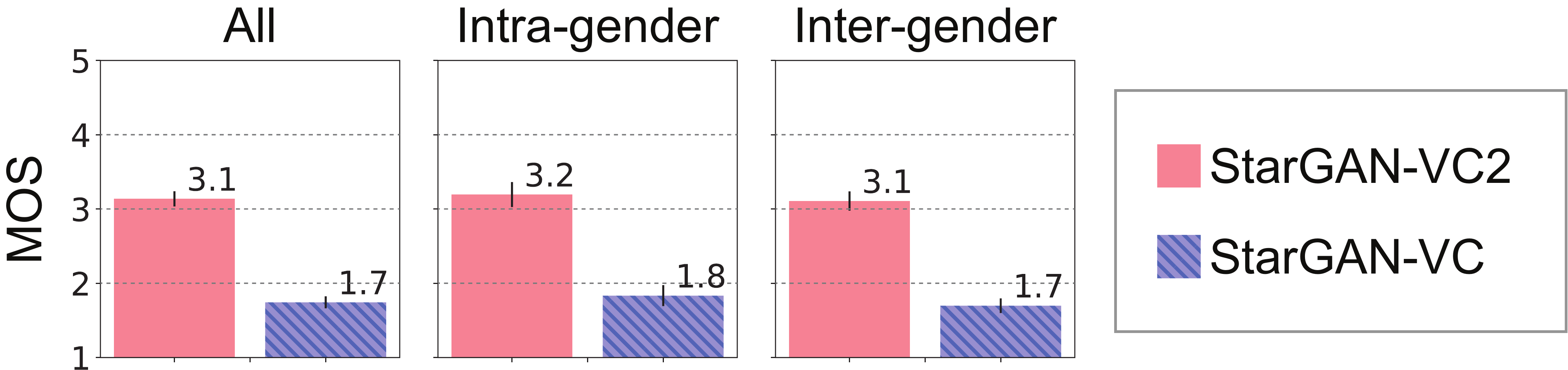}}
  \vspace{-3mm}
  \caption{MOS for naturalness with 95\% confidence intervals}
  \label{fig:mos}
  \vspace{-3mm}
\end{figure}

\begin{figure}[t]
  \centering
  \centerline{\includegraphics[width=1\columnwidth]{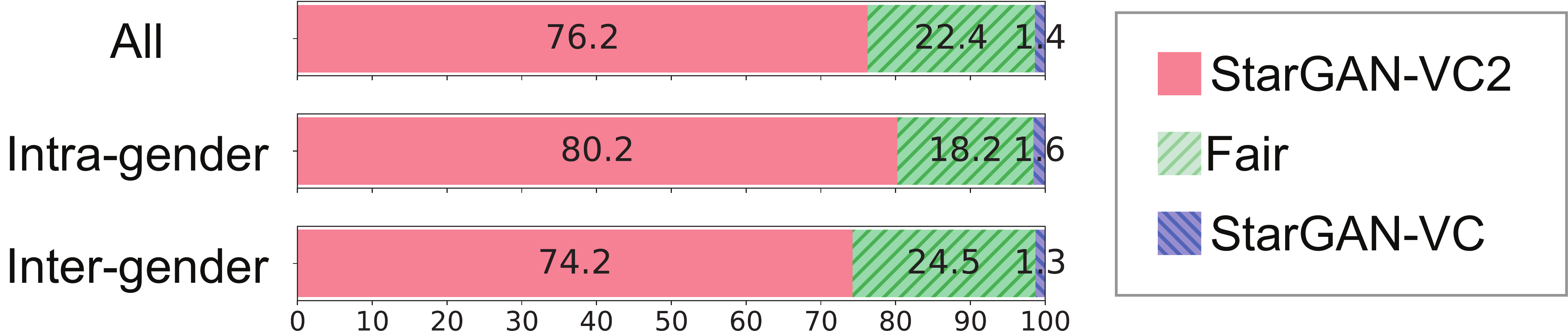}}
  \vspace{-3mm}
  \caption{Average preference scores (\%) on speaker similarity}
  \label{fig:xab}
  \vspace{-5mm}
\end{figure}

\subsection{Subjective evaluation}
\label{subsec:subjective}

We conducted listening tests to analyze the performance compared with StarGAN-VC \cite{HKameokaSLT2018}, which is a state-of-the-art multi-domain non-parallel VC. To measure naturalness, we conducted a mean opinion score (MOS) test (5: excellent to 1: bad), in which we included the analysis-synthesized speech (which is the upper limit of the converted speech) as a reference (MOS: 4.2). For each model, we generated 36 sentences ($4 \times 3$ source-target combinations $\times$ 3 sentences). We conducted an XAB test to measure speaker similarity. Here, ``X'' was target speech and ``A'' and ``B'' were speech converted by StarGAN-VC and StarGAN-VC2, respectively. For each model, we generated 24 sentences ($4 \times 3$ source-target combinations $\times$ 2 sentences). To eliminate bias in the order of stimuli, we presented all pairs in both orders (AB and BA). For each sentence pair, the listeners were asked to select their preferred one (``A'' or ``B'') or ``Fair.'' 12 well-educated English speakers participated in the tests.

Figures~\ref{fig:mos} and \ref{fig:xab} show the MOS for naturalness and the preference scores for speaker similarity, respectively. We summarized the results on the basis of three categories: all conversion, inter-gender conversion, and intra-gender conversion. These results empirically demonstrate that StarGAN-VC2 outperforms StarGAN-VC in terms of both naturalness and speaker similarity for every category.

\section{Conclusions}
\label{sec:conclusions}

To advance the research on multi-domain non-parallel VC, we have rethought conditional methods in StarGAN-VC in two aspects: training objectives and network architectures. We developed a source-and-target conditional adversarial loss for the former and a modulation-based conditional method for the latter and have proposed StarGAN-VC2 incorporating them. The empirical studies on non-parallel multi-speaker VC demonstrate that StarGAN-VC2 outperforms StarGAN-VC in both objective and subjective measures. StarGAN-VC2 is a general model for multi-domain VC and is not limited to multi-speaker VC. Adapting it to other tasks (e.g., multi-emotion VC and multi-pronunciation VC) remains a promising future direction.

\noindent
\textbf{Acknowledgements:}
This work was supported by JSPS KAKENHI 17H01763.

\clearpage
\bibliographystyle{IEEEtran}
\bibliography{refs}

\end{document}